# Trajectory Deflection of Spinning Magnetic Microparticles, the Magnus Effect at the Microscale[1]

M. Solsona [a], H. Keizer [a], H. d. Boer [a], Y.P. Klein [b], W. Olthuis [a], L. Abelmann [,c] and A. van den Berg [a]

[a] BIOS-Lab on a chip group, MESA+ Institute for Nanotechnology, Max Planck-University of Twente Center for Complex Fluid Dynamics University of Twente, Drienerlolaan 5, Enschede, 7522 NB, Netherlands
[b] Mesoscale chemical systems group, MESA+ Institute for Nanotechnology, University of Twente, Drienerlolaan 5, Enschede, 7522 NB, Netherlands
[c] KIST Europe, Saarland University, 66123, Saarbrücken, Germany

The deflection due to the Magnus force of magnetic particles with a diameter of 80 μm dropping through fluids and rotating in a magnetic field was measured. With Reynolds number for this experiment around 1, we found trajectory deflections of the order of 1 degree, in agreement within measurement error with theory. This method holds promise for the sorting and analysis of the distribution in magnetic moment and particle diameter of suspensions of microparticles, such as applied in catalysis, or objects loaded with magnetic particles.

## 1| Introduction

A rotating object moving through a medium experiences a Magnus force that is perpendicular to both the axis of rotation and the direction of motion.[1] This effect is well known in ball sports, for instance as topspin in tennis. The Magnus force is an inertial effect and therefore is most effective for large objects moving at high velocity, such as soccer balls[2] or planets forming in a protoplanetary disc.[3] In these situations, the flow is normally turbulent, characterized by much higher linear ($R_e$) and rotational Reynolds number ($R_\omega$). In this paper, we investigate the Magnus force for laminar flow conditions, at Reynolds numbers close to unity. The experiments were performed with spheres of only 80 μm in diameter, rotating less than five revolutions per second and moving at about one centimeter per second through water.

The Magnus force has technological relevance, since it can be used for magnetic separation of microparticles.[4] In contrast to magnetic separation by force gradients, Magnus separation is performed in a uniform field. The Magnus force separates primarily on particle size, since the deflection of the particles is proportional to the square of their diameter. When increasing the rotation frequency of the field, particles with low magnetic moment can no longer follow the field. So by tuning the rotation frequency, one can independently separate on magnetic moment as well.

The theory of the Magnus effect at low $R_e$ has been studied in detail.[5] It was shown that the Magnus force is linearly proportional to the rotation up to values in the order of one hundred. However, measurements of the Magnus force are not reported for values below a few hundreds.[4] A serious experimental complication is that the deflection of the trajectory of the objects approaches very low values, which complicates analysis. The solution we chose is to reduce the particle size, which also allows us to benefit from microfluidic systems.

Microfluidic technologies have been used extensively to sort cells and microparticles using forces that are a function of combinations of particle properties such as size,[6,7] shape, density,[8] permittivity, susceptibility and magnetic moment.[9–14] The use of these forces in well-controlled laminar fluids with external actuators enables important applications such as sorting of cancer cells[15] or catalyst particles.[16] In addition to forces, torques can be applied on particles by rotating fields.[17] At low $R_e$, linear and angular drag forces are balanced by the applied torque immediately, resulting in constant linear and angular velocity[18–20] and therefore a constant Magnus force.

In the experiment we present here, the force resulting in linear velocity was provided by gravitation, and the torque leading to angular velocity was provided by a rotating magnetic field acting on anisotropic magnetic Janus particles with a diameter of 80 μm. These particles are in the same size range as catalyst particles,[16,21] so that the results are of immediate technological relevance. The resulting deflection of the particles was observed by a microscope at different rotation speeds and medium viscosities, and compared to an approximate model for low $R_e$.

We show that if the experiment is performed carefully, the Magnus force can still be observed for $R_e$ around unity, and that particle trajectories can be predicted by the Rubinov and Keller model. Since there are no unexpected small-size effects, these results encourage the application of Magnus separation at the microscale using microfluidic technology.

## 2 | Theory

A sphere dropping through a fluid experiences drag forces on its surface. If the velocity is sufficiently low, the relative velocity of the fluid molecules at the surface is zero, and all drag forces are due to shear between the molecules in the fluid itself. At low velocity, the drag force on a small surface element is proportional to the relative velocity of the fluid at small distance from that surface. The velocity of the sphere reaches a maximum when the total drag force integrated over the surface is balanced by the gravitational force

If the sphere does not rotate, the fluid velocity is mirror symmetric to the vertical axis (the falling direction). All horizontal components of the drag forces compensate each other, and all vertical forces add up to a net drag force opposite to the velocity (Figure 1a). If the sphere rotates, the velocity field is modified. At very low rotation velocity, the field is mirror symmetric with respect to the horizontal plane through the center of the sphere. All horizontal components of the drag forces above the plane are compensated by opposite horizontal forces below that plane. As a result, there is no net horizontal force (Figure 1b). At higher rotation velocities however, the fluid approaching the bottom of the sphere (the front side) needs a non-negligible distance to accelerate. At the top of the sphere (the back side), the fluid needs a certain distance to decelerate. As a result, the symmetry with respect to the horizontal plane is lost. Horizontal drag force components above the horizontal plane are no longer compensated by components below that plane. Consequently, there is a net drag force component perpendicular to the vertical, and the sphere trajectory is deflected from the vertical axis (Figure 1c).

This resulting force was initially discovered by Isaac Newton and two centuries later again by H.G. Magnus.[1] In essence, it is an inertial effect[22,23] and therefore increases with increasing linear Reynolds number ($R_e= 2ur/v_f$) and rotational Reynolds number ($R_\Omega = \Omega r^2/v_f$), where $u$ [m s$^{-1}$] is the relative linear velocity between the particle and the fluid, $\Omega$ [rad s$^{-1}$] is the angular velocity (in the experimental part, we express the angular velocity in the more intuitive units of revolutions per second [rps]), $r$ [m] the radius of the particle, and $v_f$ the kinematic fluid viscosity of the fluid [m$^2$ s$^{-1}$].[24–26] Many studies have been performed in order to model this force.[27] For low Reynolds numbers, Rubinov and Keller[5] model the force as,

$$F_{\text{magnus}} = \pi \rho_f r^3 \Omega \times u, \qquad (1)$$

where $\rho_f$ [kg m$^{-3}$] is the fluid density. The force is maximum when the rotation axis is perpendicular to the relative velocity.

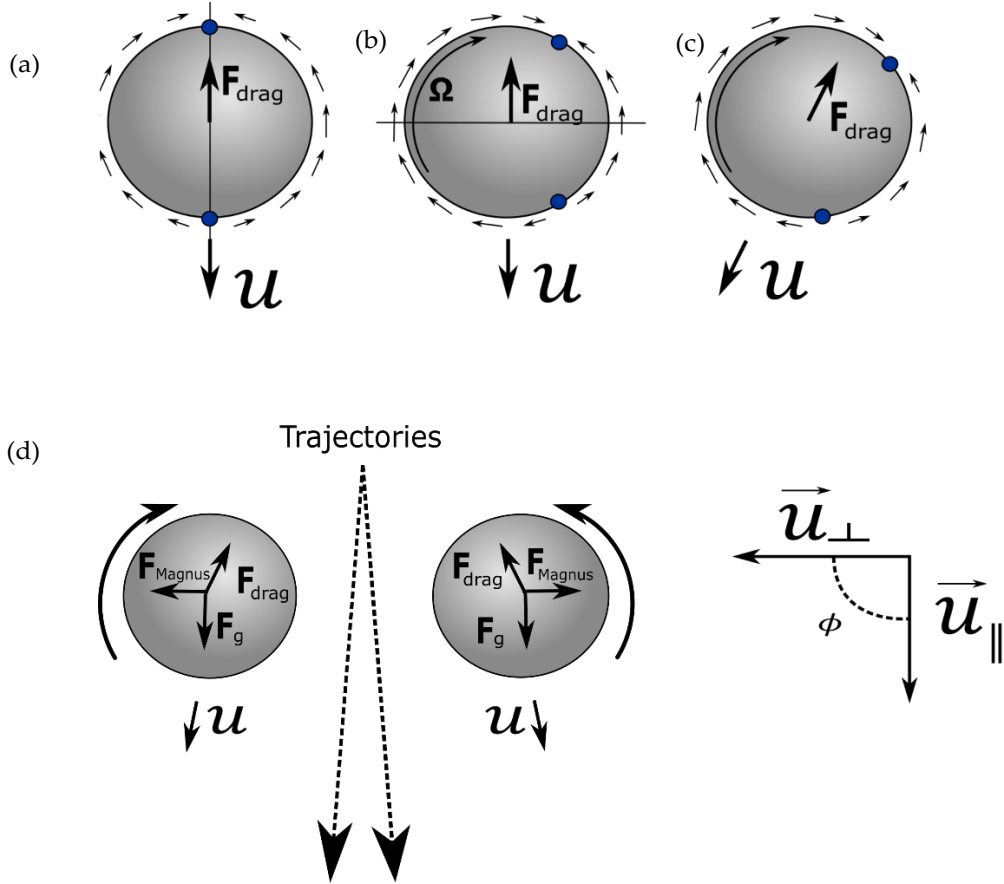

*Figure 1. Schematic drawing of the principle causing the Magnus force. A sphere dropping through a fluid experiences a drag force due to the shear force between the fluid and the sphere surface. The blue dots indicate the point where the relative fluid velocity along the surface of the sphere is zero. (a) When the sphere does not rotate, the vertical axis is an axis of mirror symmetry (indicated by the line). (b) When the sphere rotates slowly, the flow pattern shifts but remains symmetric with the horizontal plane (indicated as well). There is no net force perpendicular to the falling direction. (c) At higher rotation velocity, the fluid at the front side of the sphere needs a certain distance to accelerate, which shifts the position of zero velocity down. The resulting asymmetry in the pattern leads to a tilt in the drag force. (d) Definition of forces, particles trajectories depending on the rotation direction and velocities.*

Previous work has experimentally demonstrated the existence of the Magnus force at low $R_e$. Oesterlé et al.[28] used metal spheres a few centimeters big, attached to a thread in order to spin them, to quantify the Magnus force at $R_e$ between 10 and 140. Others[3,4,26,29] studied the phenomena at higher $R_e$, from 300 to $10^5$. To the best of our knowledge, the study of the Magnus force has never been performed at Reynolds numbers close to unity. Neither are we aware of studies with microparticles. Experiments with microparticles are challenging, since the Magnus force has a third-order dependency on the particle size (Equation 1), so the deflection decreases considerably with reduction in particle diameter.[30]

In order to rotate the particles, a rotating magnetic field was used so that we can apply a torque over a large spatial region.[17] The Magnus force is proportional to the rotation of the particles (Equation 1). The angular velocity is equal to the rotation of the magnet only if the particles can follow the magnetic field. If the magnet rotation speed is too high, the particles only wobble. To estimate the maximum rotation speed, we assume that the particle has a remanent magnetic moment $m_P$ [A m$^2$] and that the field $B$ [T] is small compared to the saturation field. Under these conditions, we can estimate the maximum torque from

$$\boldsymbol{\Gamma} \leq \boldsymbol{m}_\mathrm{p} \times \boldsymbol{B} \tag{2}$$

A sphere with radius $r$ rotating in a fluid with viscosity $\mu_f$ [Pa s] at an angular velocity $\Omega$ [rad s$^{-1}$] experiences a drag torque in the direction opposite to the rotation,[31]

$$\Gamma_\mathrm{d} = -8\pi r^3 \mu_\mathrm{f}\Omega. \tag{3}$$

By balancing the magnetic and angular drag torques, we obtain the maximum angular velocity

$$\Omega \leq \frac{m_\mathrm{p}}{r^3}\frac{B}{8\pi\mu_\mathrm{f}} \tag{4}$$

In addition to being pulled sideways by the Magnus force, the particles are pulled downward by gravity as,

$$F_\mathrm{g} = (\rho_\mathrm{p} - \rho_\mathrm{f})gV_\mathrm{p}, \tag{5}$$

where $\rho_p$ [kg m$^{-3}$] is the particle mass density, $g$ [m s$^{-2}$] is the gravitational acceleration constant, and $V_p$ [m$^3$] is the volume of the particle. A sphere moving in a fluid encounters a translational drag force in the opposite direction to the movement,

$$F_\mathrm{d} = -6\pi r\mu_\mathrm{f}u. \tag{6}$$

Balancing the forces, and considering that the magnetic torque is applied perpendicular to $\boldsymbol{g}$, we obtain the velocity of the particle, which we decompose into the translational and perpendicular components, see Figure 1:

$$u_\parallel = \frac{(\rho_\mathrm{p} - \rho_\mathrm{f})gV_\mathrm{P}}{6\pi\mu_\mathrm{f}r}, \tag{7}$$

$$u_\perp = \frac{\rho_\mathrm{f}r^3\Omega u_\parallel}{6\mu_\mathrm{f}r}. \tag{8}$$

The tilt angle of the trajectory is therefore:

$$\phi = \tan^{-1}\left(\frac{u_\perp}{u_\parallel}\right) = \tan^{-1}\left(\frac{\rho_\mathrm{f}r^2}{6\mu_\mathrm{f}}\Omega\right) = \tan^{-1}\left(\frac{r^2}{6v_\mathrm{f}}\Omega\right) \approx \frac{r^2}{6v_\mathrm{f}}\Omega \tag{9}$$

The approximation is valid for small angles. When the angular speed is zero, the particles follow the gravitational force and the tilt angle is zero. The tilt angle increases with increasing sphere radius $r$, and decreases with increasing kinematic fluid viscosity $v_f$ [m² s⁻¹].

Small particles with high magnetic moment can rotate faster than big particles with low moment (Equation 4). The tilt angle is, however, only dependent on the particle size (Equation 9). So by selecting combinations of rotating speed $\Omega$ and field strength $B$, and filtering out certain tilt angles, we have a method to discriminate particles based on radius, to some extend irrespective of magnetic moment.

Next to the Magnus force, the particles will experience a magnetic force along the magnetic field gradient,

$$F_m = m_p \nabla B \tag{10}$$

which will also lead to a tilt angle. Fortunately, since the magnetic moment will align with the field direction, the direction of the gradient is independent of the sign of the field. The sign of the tilt angle, however, is determined by the rotation direction. So by measuring both rotations, any tilt due to a magnetic force gradient can be subtracted.

## 3 | Materials and Methods

There are three requirements that the fluidic system must meet. First, due to the small deviations expected by the Magnus force and in order to facilitate the following trajectory measurement, the particles should start at very similar positions. Second, the system should be long enough to track long trajectories, and third, the particles should be clearly visible through the fluidic system walls. Figure 2 shows the 3-D printed fluidic system with thee inlets and one outlet. Two of the three inlets are used to introduce the liquid, and the third inlet is used to introduce the magnetic particles, also see Figure S1a in Supporting Information. The chamber is 7 cm long and 0.5 cm wide and deep. In order to observe the particles and seal the chamber, a glass slide was glued on the front part of the chip, see Figure S1a in Supporting Information. To maximize the deflection caused by the Magnus force, the time of the particles inside the chamber should be increased. Therefore, the system will be used with no-flow conditions, letting the particles sink from the inlet to the outlet of the chamber. A very similar particle starting point inside the fluidic system is crucial due to the small deflection expected caused by the Magnus force. Normally, liquid flow or magnetic, electric or acoustic fields are used to focus the particles inside a microfluidic channel. However, due to the no-flow conditions and the trouble of introducing actuators inside the 3-D printed chip, a new method to focus the particles at the same position was developed. Figure 2c shows a cut of the particle's inlet of the microfluidic chip where a zig-zag inlet can be seen. As can be seen in Figure S2 in Supplementary Information, particles were rolling downward. Although some of the particles started at very similar positions, this varied due to its large sensitivity on other factors such as: the amount of particles arriving to the big chamber at the same time and disturbing each other or by any small flow perturbation caused by the pipetting of the particles inside the inlet.

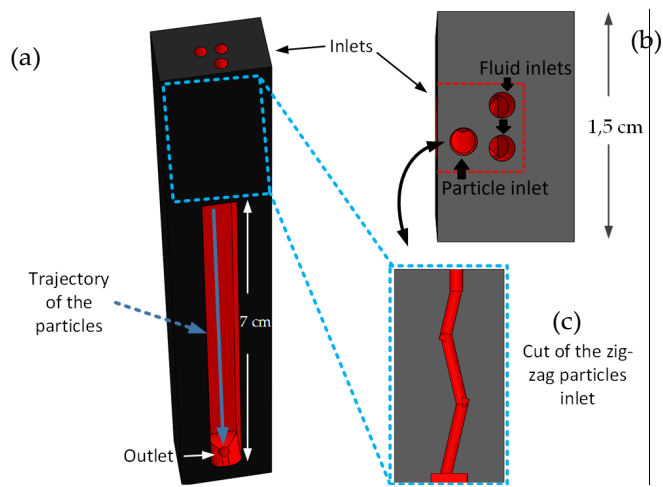

*Figure 2. (a) Schematic drawing of the fluidic chip showing the three inlets and chamber where the particles rotate, (b) the three inlets and (c) cut of the zig-zag particles inlet.*

In order to rotate the particles and avoid any attraction to the magnets, the magnetic field that is used should be as strong and homogeneous as possible. Therefore, bigger magnets providing less gradient over the observation area are better than smaller magnets. Also, by placing another magnet on the opposite side of the fluidic chip, the gradient, and therefore its magnetic force toward the magnets, will be reduced, see Figure 3. Eight N42 magnets, 1 cm wide, 4 cm deep, 1 cm long and a magnetization of 1.3 T along the longest side, were purchased from Supermagnete.[32] Each rotating arm consisted of four magnets, as can be seen in Figure S3b. The field strength in the center between the magnets was 122 ± 5 mT. In order to observe the particle's trajectory, a silicon wafer acting as a mirror was glued at a 45° angle to the fluidic system's wall, see Figure S1b and c in Supplementary Information.

Both magnets should rotate at the same speed in order to avoid magnetic field distortion, but due to the fluidic connections and tubing, both magnets could not be attached together. Therefore, a new mechanical system was developed to rotate both magnets at the same speed. Figure S3a in Supporting Information shows the system consisting of an electrical motor (Crouzet DC motor, model 820580002) and a gear box with a reduction ratio of 3.4, see Figure S1b and c in Supplementary Information. Also, for security reasons, a new system was developed to mount the magnets in the rotating arms. As can be seen in Figure S3b in Supporting Information, first the magnets were stuck together with a separator in the middle and subsequently attached together to one of the rotator's arms. Thereafter, the other rotating arm was brought closer and connected to the second magnet's support, which allows safe separation of both magnets. The magnets repelled each other due to their configuration, therefore the second arm was pushed away by the magnets. The separation of the magnets can be adjusted with extra screws in the rotating arms.

The rotating speed of the magnets was adjusted, controlled by a Conrad PS 405 Pro power supply. The rotation speed was calibrated with a stroboscope (LED-stroboscope HELIO-STROB micro2). A Grasshopper3 GS3-U3-23S6M high-speed camera was used to observe and record the trajectories of the particles. The 3-D printed chip was designed in SolidWorks and printed with a Formlabs Form 2 printer.

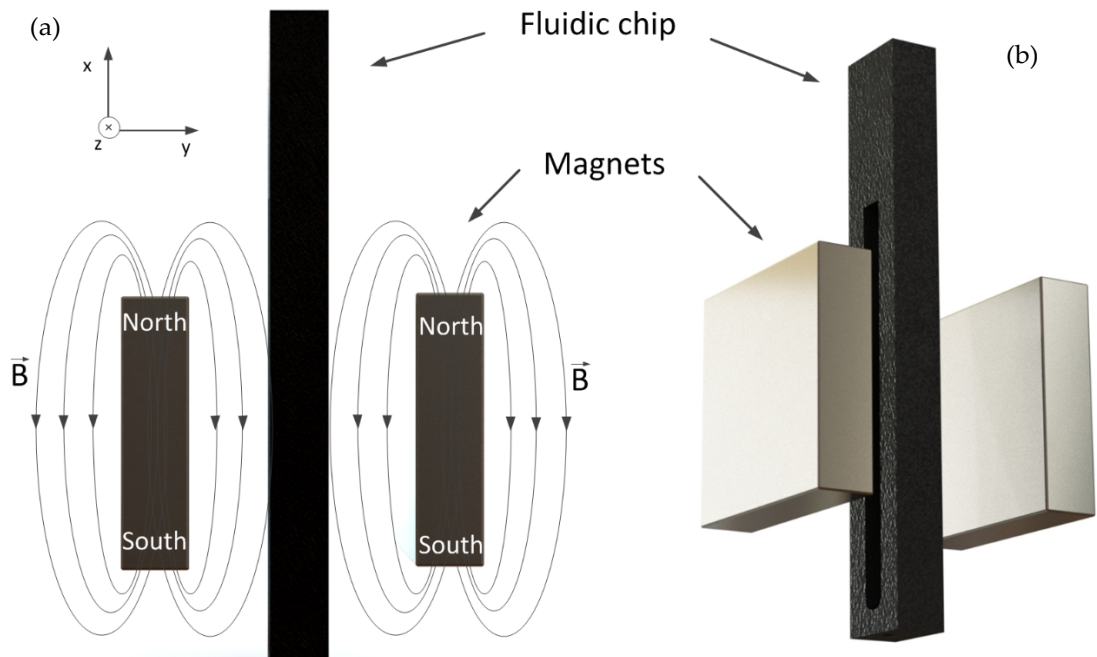

(a)

x, z, y (coordinate axes)

Fluidic chip

Magnets

North

South

$\vec{B}$

North

South

$\vec{B}$

(b)

*Figure 3. (a) Schematic (side view) drawing and magnetic field distribution used to rotate the Janus particles inside the fluidic chip. (b) Diagonal view of the magnets' configuration.*

The particles used in this study are magnetic Janus particles, 70–90 μm in diameter, purchased from Cospheric.[33] Their core is made of borosilicate, and they are half-coated with a superparamagnetic material.

Five hundred and forty-three Janus particles were placed between two 5×5 mm tape layers and introduced into a Quantum Design PPMS-VSM (physical property measurement system vibrating sample magnetometer) to measure their magnetic moment. The scan rate was 1.5 [mT s⁻¹] from 5 T to −5 T and up to 5 T again at room temperature. Figure S4 in Supplementary Information presents the experimentally obtained magnetization of the Janus particles. From the magnetization curve, it can be concluded that the Janus particles have hysteresis and that their magnetic moment saturates at ≈ 0.5 T, having a susceptibility of 0.06 [nA m² T⁻¹] and a remanent moment $m_P$ of 1.5×10⁻¹¹ [A m²] per particle.

The resulting magnetic force exerted by the magnet on the particles (Equation 10) was estimated from the magnet geometry using analytical integration (Cades[34]), see section 5 in SI. We are mostly concerned with variations in the magnetic force for different particle trajectories. When the particles fall exactly along the center between the magnets, the force and force gradient is zero. If particles rotating both directions start falling at the same point, a constant force will merely cause an offset that is cancelled by measuring the difference between clockwise and anti-clockwise rotation. However, in each experiment, particles rotating both sides had a small difference in position close to 1 mm due to the inaccuracy of the focussing system. Magnetic forces in z-direction depending on z-position (z = 0.1, 1 and 5 mm) range from 0.1 to 35 pN, see Figure S5 a. Even particles flowing 2 mm away from the magnetic field center (x,y,z)=(0,0,0) have a force difference close to 10 pN, see Figure S5 b. The gradient of the magnetic force in the z-direction, so in the same direction as the Magnus force, was estimated to be 7.5$z_0$ nN/m, see Figure S6, where $z_0$ is the off-center position. Therefore, particles rotating 1 mm apart had a difference in force of 7.5 pN. The magnetic attraction by the magnets is therefore of the same magnitude as the Magnus force itself, which is in the order of 6 pN (DI water with $\mu_f$ = 1 mPa·s and 5 rps). This suggests that the deflection caused by the Magnus effect is enhanced by the magnetic force. Typical deflection differences due to the Magnus effect are less than 0.1 mm, resulting in maximum magnetic force differences of 0.75 pN.

Two solutions were used, DI water and a mix of glycerol and water in a 1:3 volume ratio. Three pipette tips were glued in the three inlets to facilitate the introduction of liquids and particles. The chip was filled with a pipette until the liquid reached the top of the pipette tips. Matlab was used to track the particle trajectories, and Cades was used to simulate the magnetic field and magnetic force.

Three different experiments were performed where just the liquid viscosity and the rotation speed of the particles were modified. The experimental procedure consisted of different steps. First, the rotation of the magnets was set to a given angular speed. Second, the Janus particles were introduced in the fluidic chip via a pipette. Thereafter, they rolled and fell into the main channel at $u \approx 0.01\ m/s$. Third, their trajectories over 1 cm length inside the fluidic system were recorded and subsequently analyzed. Six different steps in each experiment were performed, three clockwise and three anti-clockwise . The center of the trajectories was manually centered at 0 in order to visualize the difference between experiments.

## 4 | Results and Discussion

Figure 4 shows the experimentally obtained angles of the trajectories of the particles and their cumulative distribution function (CDF) assuming normal distributions. In Figure 4a, a clear difference can be observed between particle trajectories when the magnet is rotating either clockwise or anti-clockwise, when we used water ($v_f$ = 1.0×10⁻⁶ m²/s), and 5 rps as angular speed, meaning $R_e \approx 1$ and $R_o$=0.2. The deflection of the trajectories agrees with the theory: particles rotating clockwise moved to the left, and particles rotating anti-clockwise moved to the right. The experimental difference between the mean deflection angles for the two rotation directions was $1.2 \pm 0.3°$. This is within the error margin in agreement with the calculated angle difference of 1.2° (Equation 9). Figure 4b shows the same experiment but using 40% of the angular speed (2 rps). The measured angle difference ($0.42 \pm 0.15°$) again agrees with the calculated angle of 0.49°. When we used a liquid with more than twice the viscosity ($v_f$ = 2.3×10⁻⁶ m²/s), meaning a $R_e \approx 0.2$ and $R_o$=0.03, and low angular speed( 2 rps), the difference between particles rotating in both directions drops below the measurement error. This was an expected result since the calculated angle dropped until 0.21°. When we increased the angular velocity to 5 rps, most of the particles were no longer able to follow the magnetic field rotation. Assuming the magnetic field to be in the order of 122 mT, the maximum rotation velocity in the high-viscosity medium (μ = 2.3 mPas) is, according to Equation 2, in the order of 78 rps, so we should have observed the particles completely following the magnetic field and a deflection similar to 0.52°. We speculate that most of the particles didn not have enough magnetic material and therefore not enough torque to follow the magnetic field.

In order to obtain larger deflections and more feasible sorting systems, larger rotation speeds are needed, which can be obtained by increasing the magnetic field strength inside the fluidic system by placing the magnets closer together). It should be noted that the Magnus force is smaller than typical magnetic forces on microparticles in microfluidic devices. Comparing different systems using magnetic fields to sort particles it can be observed that similar forces are accomplished (1–35 pN) however with smaller particles (1-7.3 μm). [35-40] The experimental system was constructed for minimal magnetic field gradients, and therefore required big magnets. The size can be optimized however, especially if we allow magnetic field gradients on top of the Magnus force for additional sorting

## 5 | Conclusion

In conclusion, we made magnetic Janus particles of 80 μm diameter fall through a liquid with varying viscosity in the presence of a rotating magnetic field with varying angular velocity. The Reynolds number for the particle movement was ≈ 1. Below a threshold field, the particles rotated with the field. These rotating particles were therefore subject to a Magnus force that caused a measurable tilt of up to 1.2° in their trajectories. The direction of the tilt agrees with rotation direction and with a simple model based on the Rubinov and Keller approximation for the Magnus force at low Reynolds numbers combined with linear viscous drag. The tilt angle increased with increasing rotation velocity and decreasing kinematic viscosity. The minimum field value for rotation increases with increasing rotation velocity and viscosity of the medium. Most of the particles no longer followed a magnetic field of 122 mT in a medium with a kinematic viscosity of $2.3 \times 10^{-6}$ m²/s at a rotation velocity of 5 rps, and therefore no longer showed a tilt.

The experiments clearly demonstrate that the Magnus force on particles with a diameter of tens of micrometers is measureable. The method allows for separation of particles based on the ratio between their magnetic moment and radius to the third power.

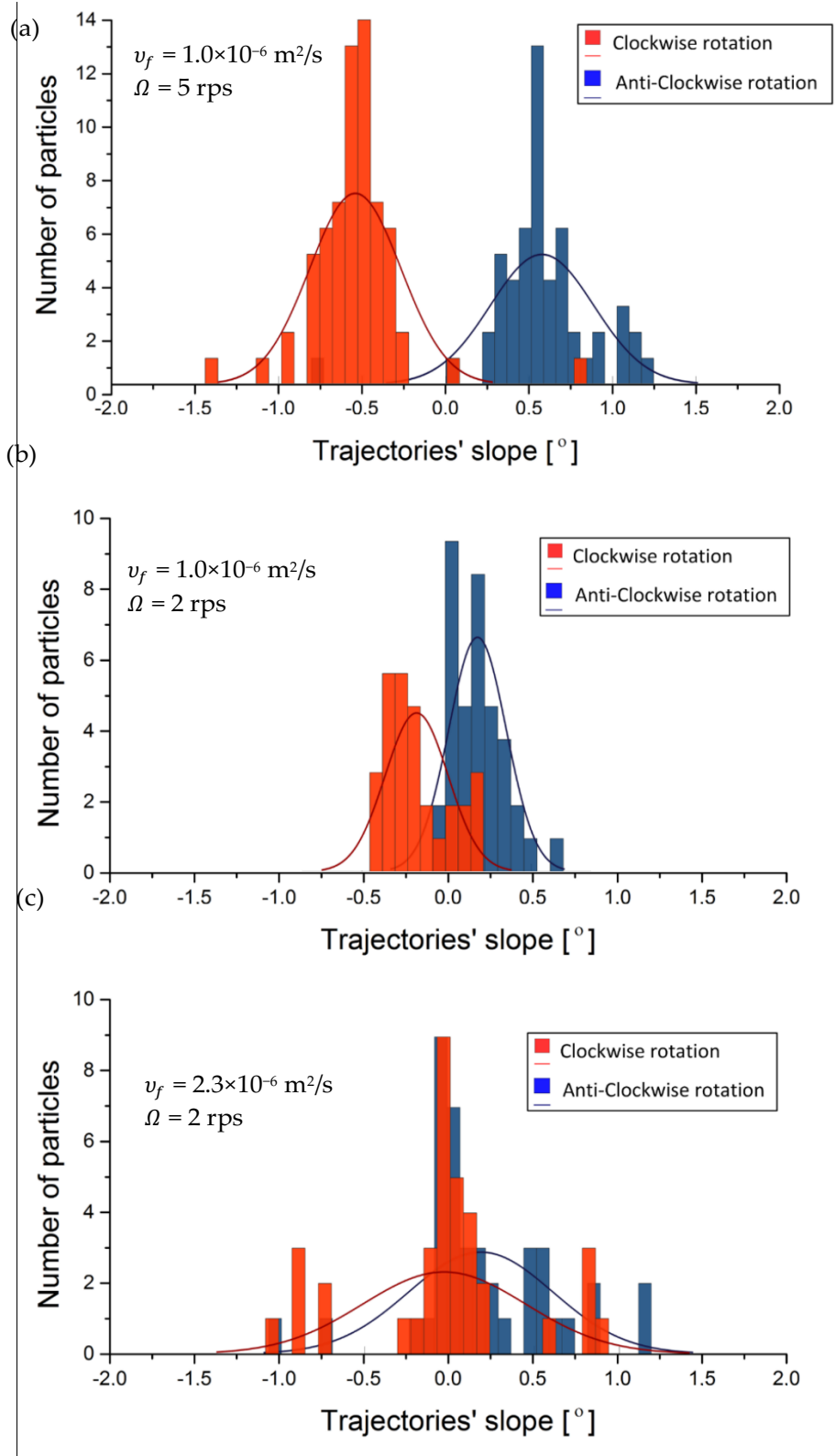

*Figure 4. Histograms and their cumulative distribution function (CDF) of the trajectory slope angles in z-direction of particles rotating clockwise (red) and anti-clockwise (blue) (a) with low viscosity (water) and 5 rps, (b) low viscosity (water) and 2 rps and (c) high viscosity (mix of water and glycerol 3/1 v/v) and 2 rps.*

## 6 | References

# Supporting Information

**Section S1: General setup**

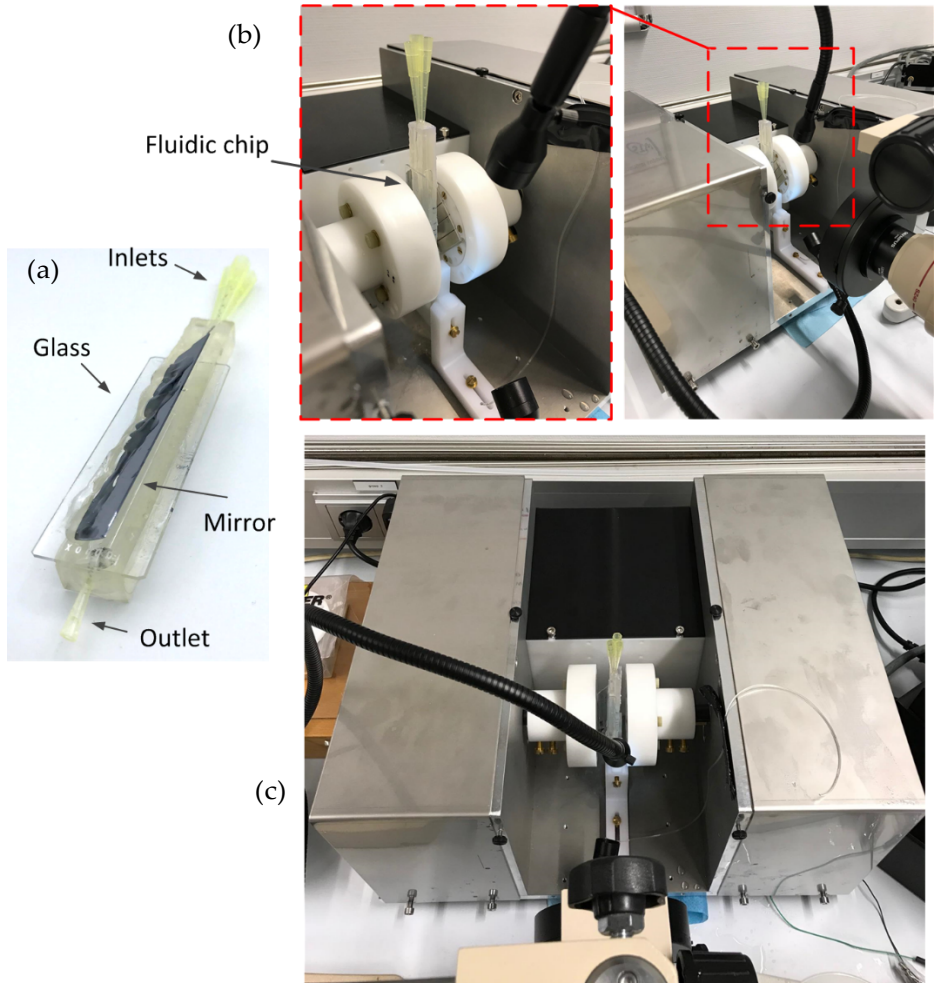

*Figure S1. (a) Image of the fluidic system with a glass to seal the chip and observe the janus particles, a mirror to reflect the trajectories of the janus particles and 3 pipette tips to introduce the fluid and particles. (b) Images of the setup where both magnets are placed at both sides of the fluidic chip. (c) Top-view of the setup.*

**Section S2: Microfluidic inlet**

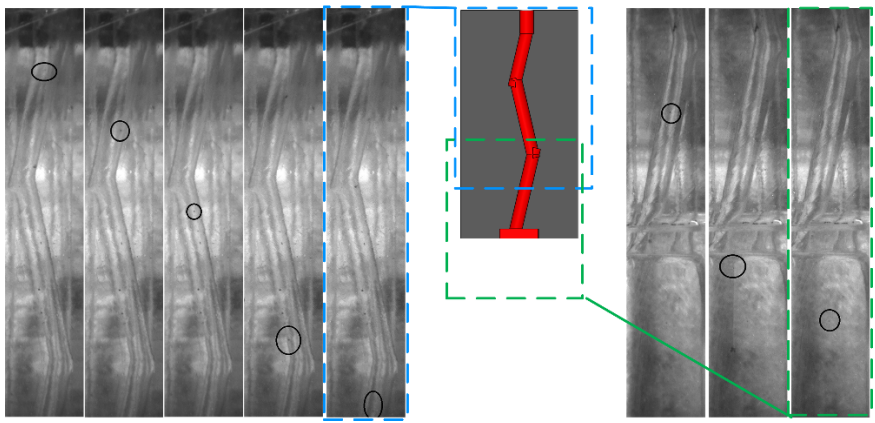

*Figure S2. Zig-zag inlet fabricated to introduce the janus particles at similar starting positions. Inside the black circle a particle rolling down can be followed until entering in the main channel.*

**Section S3: Sketch of the rotating magnetic field setup**

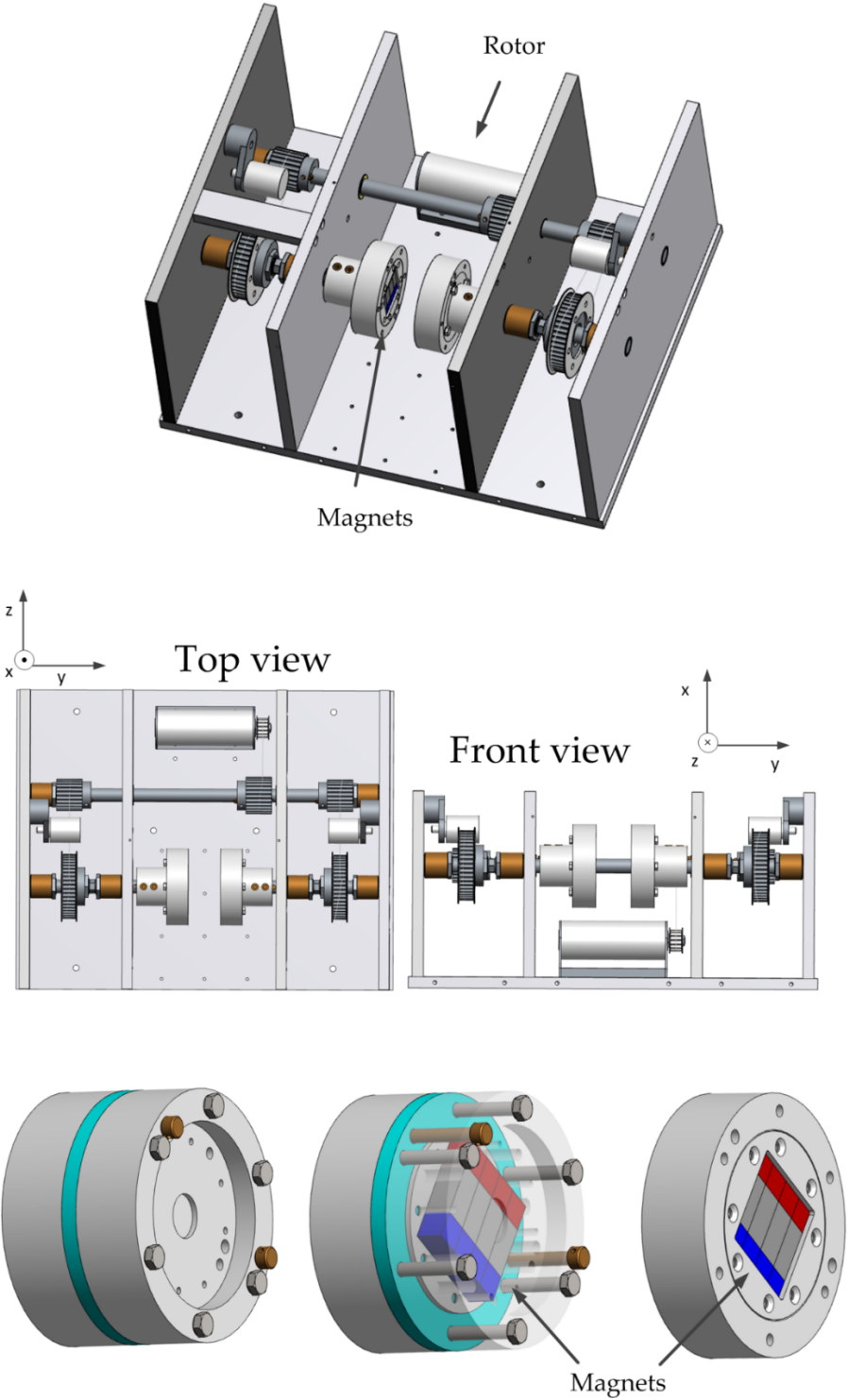

*Figure S3. Solidworks drawing of the (a) rotating magnetic field setup where the electric motor rotates the magnets at the same rotational velocity and (b) the mounting of magnets setup.*

**Section S4: Magnetic moment of the janus microparticles**

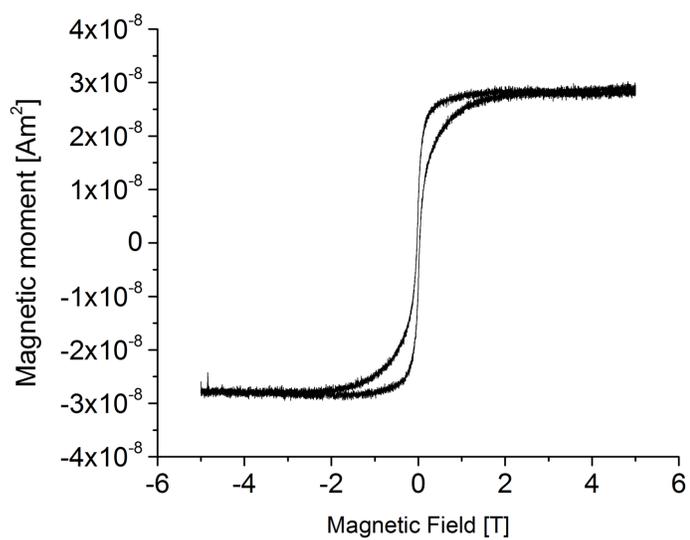

*Figure S4. Magnetic moment of 453 janus particles from -5 to 5 T.*

## Section S5: Simulation of Magnetic field and force.

The force was simulated using Cades considering a paramagnetic sphere with a radius of 0.0169 mm (equal the volume of the magnetic material per particle) and a susceptibility of -0.016 taken from Figure S4.

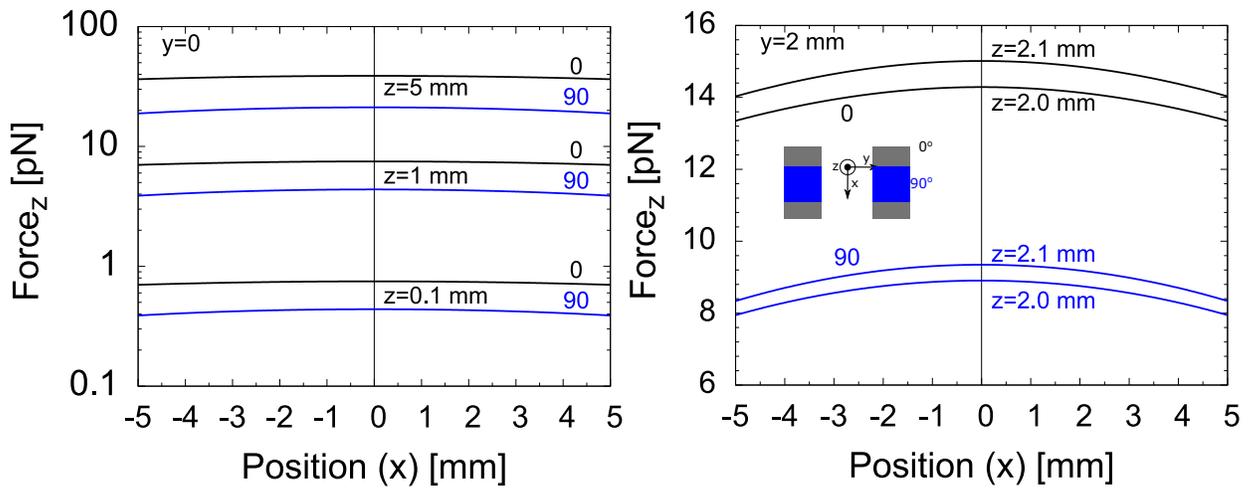

*Figure S5. Magnetic force in z direction (same direction as the Magnus effect) across the x direction (particle trajectory) for varying offset in the z direction in the center between the magnets (left), and at at an offset y of 2 mm (right), The blue curves are related to the same forces when the magnets a rotated 90°.*

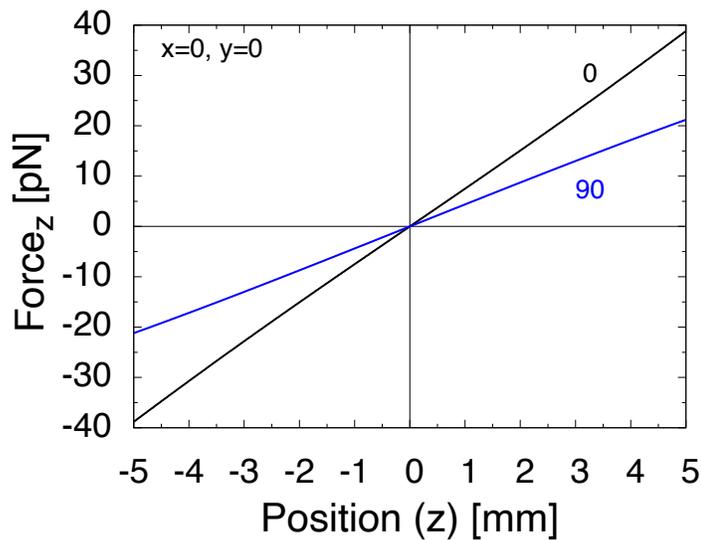

*Figure S6. Force in the z-direction (same direction as the Magnus effect) as a function of the offset in the horizontal plane (z-direction), at x=0 and y=0, for magnets in rest position and rotated by 90° (blue curve).*